\documentclass[11pt,a4paper,fleqn]{article}
\usepackage{amssymb}
\title{\Large{\textbf{On the Massless Vector Fields in a Rindler Space}}}
\author{Roberto Soldati\\
\emph{Istituto Nazionale di Fisica Nucleare - Sezione di Bologna}\\
\textit{Dipartimento di Fisica - Universit\`a di Bologna}\\
\\ \\
Caterina Specchia\\
\emph{ETH Z\"{u}rich, Institut f\"{u}r Theoretische Physik}}
\begin{document}
\maketitle
\begin{abstract}
We study the quantum theory of the mass-less vector fields on the Rindler space. 
We evaluate the Bogoliubov coefficients by means of a new technique based upon the use
of light-front coordinates and Mellin transform.
We briefly comment about the ensuing Unruh effect
and its consequences.
\end{abstract}
%
%
\subsection*{Introduction}
The production, propagation and detection of real photons in a non-inertial reference frame is a 
highly nontrivial subject \cite{birreldavies,hawton}. 
The matter is that, if one aims to describe those features in a coordinate 
independent way, i.e. local diffeomorphism invariant in the physical \textit{3+1} dimensional
space-time, one has to properly separate the physical and nonphysical polarization modes, the
latter ones being necessarily present in any gauge and diffeomorphism invariant general formulation
of the quantum theory:
this becomes a hard task which has not yet been reached, even for a  uniformly accelerated 
frame in a flat space-time referred to a curved coordinate system \cite{CSthesis}.
Actually, in the nearly whole Literature on the non-inertial effects on quantum fields, 
just like the celebrated Unruh effect \cite{unruh}, or about quantum field theory in curved spaces,
the emphasis, examples and applications are always centered around the real scalar field case
\cite{unruh,fulling,fedotov,lonsol11,castorina,cotaescu} up to some few exceptions concerning the realistic electromagnetic 
or Proca vector fields \cite{linet,lenz,CHM08} or the Dirac and Majorana spinor fields \cite{oriti,lonsol13}.
Recently, a quite interesting attempt to investigate, both from a theoretical and operational points of view, the process
of emission, propagation and detection of electromagnetic radiation in a uniformly  accelerated 
reference frame has been pursued in \cite{hawton} in the four dimensional
space-time. However, to the aim of avoiding the difficult disentanglement
of the physical and nonphysical polarization, the Author does restrict the dynamics
along the acceleration direction, just turning to a 2-dimensional setting and to the treatment of the 
radiation field in terms of a conventional mass-less scalar degree of freedom in a \textit{1+1}
dimensional Minkowskian space $ \mathfrak{M}_{1,1} $. But in so doing, it turns out that the correct
canonical quantization \cite{BGLN} of the radiation field is not properly taken into account so that, consequently,
some fictitious singularities appear, which are an artifact of the 
too drastic simplification of the radiation dynamics in an accelerated reference frame hitherto employed.

It is the aim of the present short note to fill this lack and to provide a fully consistent Lorentz and gauge
invariant quantum theory for the lineal, i.e. on a one dimensional spatial real line, radiation field. In so doing,
thanks to a new method based upon the use of the Mellin transforms and the light-front system of coordinates,
which has been recently developed by Aref'eva and Volovich \cite{irina}, the
Bogoliubov coefficients connecting inertial and accelerated reference frames are 
eventually and correctly obtained to be singularity free. Thanks to the present contribution, the quite relevant
and interesting operational analysis developed in \cite{hawton} is actually set on a firm and reliable
framework and might become seminal for some future experimental verification.

\subsection*{Rindler space and coordinate system}
In this Section we aim to briefly collect and recall the most useful systems
of coordinates, which are
utmost suitable to describe the field dynamics for a uniformly accelerated Observer in the Rindler space
\cite{rindler},
a non-inertial flat space-time in two dimensional curved coordinate systems -- many more details and
useful properties, aspects and relevant features can be found in \cite{sokolovsky}. 
To this concern, consider a two dimensional Minkowskian space  $\mathfrak{M}_{1,1}$ with coordinates
\footnote{Throughout this note we shall use natural units 
$ \hbar=c=1 $ unless explicitly stated.}
\[\mathrm x^{\,\mu}=(x^0,x^1)=(t,x)\]  
and metric
\begin{equation}
\mathrm ds^2=g_{\mu\nu}\,\mathrm dx^{\,\mu}\,\mathrm dx^{\,\nu}=
\mathrm dt^2-\mathrm dx^2=\mathrm dx^-\,\mathrm dx^+
\label{2metrica} 
\end{equation}
where $ g_{\mu\nu}=g^{\,\mu\nu} = {\rm diag}\,(+,-)$ is the metric tensor while
\[x^-=t-x\qquad\quad x^+=t+x\]
are the standard light-front coordinates.
If we perform the non-inertial coordinate transformation in the restricted space-like region 
$ x>|t| $ of the Minkowskian space
\begin{eqnarray}
\left\lbrace 
\begin{array}{c}
\mathrm a\,t=\exp\{\mathrm a\,\xi\}\,\sinh(\mathrm a\eta)\\
\mathrm a\,x=\exp\{\mathrm a\,\xi\}\,\cosh(\mathrm a\eta)
\end{array}
\right. \qquad\quad x>|t|
\end{eqnarray}
with $\rm a>0$ and $ \eta\,,\,\xi\in\mathbb R\,, $ or equivalently
\begin{eqnarray}
\left\lbrace 
\begin{array}{c}
u=\mathrm a^{-1} \exp \{-\,\mathrm a\,(\eta-\xi)\}=-\,x^-\\
v=\mathrm a^{-1} \exp \{\,\mathrm a\,(\eta+\xi)\}=x^+
\end{array}
\right. 
\label{uv} 
\end{eqnarray}
 then the line element of equation (\ref{2metrica}) becomes
\begin{equation}
\mathrm ds^2
=\mathrm e^{\,2\mathrm a\xi}\left(\mathrm d\eta^2-\mathrm d\xi^2\right)
\label{2metrica'} 
\end{equation}
Moreover we can readily obtain the inversion formul\ae\ for $ x>|t| $
\begin{eqnarray}
&&\mathrm a\,\xi=\ln\mathrm a\sqrt{uv}
=\ln\mathrm a\sqrt{x^2-t^2}
\\
&&\mathrm a\,\eta=\ln\sqrt{\frac{v}{u}}
=\ln\sqrt{\frac{x+t}{x-t}}
\end{eqnarray}
whence we get
\begin{eqnarray}
\frac{\partial\,\eta}{\partial\,t}=\frac{\partial\,\xi}{\partial\,x}
=\frac{x}{\rm a}\cdot\frac{1}{x^2-t^2}
=\exp\{-\,\mathrm a\,\xi\}\,\cosh(\mathrm a\eta)
\\
\frac{\partial\,\eta}{\partial\,x}=\frac{\partial\,\xi}{\partial\,t}
=\left( -\,\frac{t}{\rm a}\right) \frac{1}{x^2-t^2}
=-\,\exp\{-\,\mathrm a\,\xi\}\,\sinh(\mathrm a\eta)
\end{eqnarray}
in such a manner that if we set
$x^{\,\prime\mu}=(\eta,\xi)$
then we find
\begin{eqnarray}
\left(\frac{\partial x^{\,\prime\mu}}{\partial x^{\,\nu}}\right) &=&
\frac{1}{x^2-t^2}\left\lgroup
\begin{array}{cc}
x & -\,t\\-\,t &x
\end{array}
\right\rgroup\frac{1}{\rm a}\qquad\quad(\,x>|t|\,)\\
&=&\exp\{-\,\mathrm a\,\xi\}
\left\lgroup
\begin{array}{cc}
\cosh(\mathrm a\eta) & -\,\sinh(\mathrm a\eta)\\
-\,\sinh(\mathrm a\eta) & \cosh(\mathrm a\eta)
\end{array}
\right\rgroup
\end{eqnarray}
Finally we can readily find the transformation formul\ae\
for the differential operators, viz.
\begin{eqnarray}
\partial_-=-\,\frac{\partial\,\eta}{\partial\,u}\,\partial_{\eta}
-\,\frac{\partial\,\xi}{\partial\,u}\,\partial_{\xi}=
\exp \{\,\mathrm a\,(\eta-\xi)\}\,\left( \partial_{\eta}-\partial_{\xi}\right) \\
\partial_+=\frac{\partial\,\eta}{\partial\,v}\,\partial_{\eta}
+\frac{\partial\,\xi}{\partial\,v}\,\partial_{\xi}=
\exp \{-\,\mathrm a\,(\eta+\xi)\}\,\left( \partial_{\eta}+\partial_{\xi}\right)\\
\square=4\partial_-\partial_+=4\mathrm e^{-\,2\mathrm a\xi}\left( \partial_{\eta}^2-\partial_{\xi}^2\right)
\end{eqnarray}
The curved coordinates $ (\eta,\xi) $ cover only a quadrant of the Minkowskian space $\mathfrak M_{1,1}\,,$
i.e. the Rindler region R named the right Rindler wedge \cite{rindler}
\[\mathrm R=\{(t,x)\in\mathfrak M_{1+1}\,|\, x>|t|\}\]
Since $t/x=\tanh (\mathrm a\eta)\,,$
lines of constant $ \eta $ are straight while lines of constant $ \xi $ are just hyperbolas
\begin{equation}
x^2-t^2=\mathrm a^{-2}\mathrm e^{\,2\mathrm a\xi}=\rm constant
\end{equation}
the asymptotic of which are the null rays $ x^\pm=0$
or $ u=\infty\,,\,v=-\infty\,. $ Thus the accelerated Observers do indefinitely 
approach the speed of light for $ \eta\rightarrow\pm\infty\,, $
while the proper time $ \bar\tau $ and the proper acceleration $ \bar{\mathrm a} $
for the accelerated Observers are respectively given by
\begin{equation}
\bar{\tau}=\eta\,\mathrm e^{\,\mathrm a\xi}\qquad\quad
\bar{\mathrm a}=\mathrm a\,\mathrm e^{\,-\,\mathrm a\xi}
\end{equation}
in such a manner that hyperbol\ae\ of large negative $ \xi\,,$ i.e. close to the
Rindler horizon $ x=t=0\,, $ do represent strongly accelerated Observers 
with a short proper time $ \bar{\tau} $.

\bigskip\noindent
As a further quite useful example of curved coordinate system to label the 
two dimensional Rindler space, consider once again the sub-spaces 
$ \rm R $ and $ \mathrm{L} $ of the Minkowskian space
$ \mathfrak{M}_{1,1} $ in the curved coordinate system 
$ (\varrho,\eta) $ associated to the uniformly accelerated Observer: namely, 
\begin{eqnarray}
\left\lbrace 
\begin{array}{c}
t=\varrho\sinh\mathrm a\eta\\
x=\pm\,\varrho\cosh\mathrm a\eta
\end{array}
\right. 
\qquad\qquad(\,\mathrm a>0\,,\,\varrho>0\,,\,\eta\in\mathbb R\,)
\end{eqnarray}
where the plus and minus signs refer to the right and left 
Rindler's wedges respectively, while the line element takes the forms
\begin{equation}
\mathrm ds^2=\mathrm dt^2-\mathrm dx^2=
\varrho^{2}\mathrm a^{2}\mathrm d\eta^{2}-\mathrm d\varrho^{\,2}
\end{equation}
whence we obtain the metric tensors
\begin{equation}
\bar g_{\alpha\beta}=
\left\lgroup
\begin{array}{cc}
1&0\\0&-1
\end{array}
\right\rgroup\qquad\quad
{g}_{\mu\nu}(\varrho)=
\left\lgroup
\begin{array}{cc}
\mathrm a^{2}\varrho^{\,2}&0\\0&-1
\end{array}
\right\rgroup
\end{equation}
so that
\begin{equation}
 g={\rm det}\, g_{\mu\nu}(\varrho)=[\,{\rm det}\, g^{\,\mu\nu}(\varrho)\,]^{-1}=
-\,\mathrm a^{2}\varrho^{\,2}
\end{equation}
Moreover, after setting
\[\bar\mathrm x^{\,\alpha}=(t,x)\qquad\quad\mathrm x^{\,\nu}=(\eta,\varrho)
\qquad\quad\alpha\,,\,\nu=0,1\]
we readily get the transformation matrix
\begin{eqnarray}
\left( \frac{\partial\,\mathrm x^{\,\nu}}{\partial\,\bar\mathrm x^{\,\alpha}}\right) =
\frac{1}{\sigma}
\left\lgroup
\begin{array}{cc}
\cosh\tau&-\sinh\tau\\
-\,\sigma\sinh\tau&\sigma\cosh\tau
\end{array}
\right\rgroup
\end{eqnarray}
where $ \tau\equiv\eta\mathrm a\,,\,\sigma\equiv\varrho\mathrm a\,. $
Notice that, in general, the transformation matrix connecting the inertial and
the accelerated reference frames can be written as the product of a local
Lorentz transformation and the Zwei-Beine field
\begin{eqnarray}
\left( \frac{\partial\,\mathrm x^{\,\nu}}{\partial\,\bar\mathrm x^{\,\alpha}}\right) =
\Lambda^{\beta}_{\phantom{\beta}\alpha}(\mathrm x)\,
\mathrm X^{\,\nu}_{\beta}(\mathrm x)
\qquad\quad
\Lambda\in O(1,1)
\label{zweibein_fields} 
\end{eqnarray}
and specifically, for the present case of a uniformly accelerated Observer,
\begin{equation}
\Lambda^{0}_{\phantom{0}0}=\Lambda^{1}_{\phantom{1}1}=\cosh\tau
\qquad\quad
\Lambda^{0}_{\phantom{0}1}=\Lambda^{1}_{\phantom{1}0}=\,-\,\sinh\tau
\end{equation}
\begin{equation}
\Lambda(\tau)=\left\lgroup
\begin{array}{cc}
\cosh\tau&-\sinh\tau\\
-\sinh\tau&\cosh\tau
\end{array}
\right\rgroup
\end{equation}
\begin{equation}
\mathrm X^{\,\nu}_{\,0}(\sigma)=\sigma^{\,-1}\,\delta^{\,\nu}_{\,0}
\qquad\quad
\mathrm X^{\,\nu}_{\,1}=\delta^{\,\nu}_{\,1}
\end{equation}
Hence we can eventually write the following chain equality
\begin{eqnarray*}
\mathrm ds^2=g_{\mu\nu}(\varrho)\,
\mathrm{dx}^{\,\mu}\,\mathrm{dx}^{\,\nu}
=\bar g^{\,\alpha\beta}\,\mathrm X^{\mu}_{\alpha}(\sigma)\,
\mathrm X^{\nu}_{\beta}(\sigma)\,\mathrm{dx}^{\,\mu}\,\mathrm{dx}^{\,\nu}
=\bar g_{\alpha\beta}\,\mathrm d\bar\mathrm x^{\,\alpha}\mathrm d\bar\mathrm x^{\,\beta}
\end{eqnarray*}
For any contravariant vector field in the Minkowski space we can derive the
corresponding vector field in the Rindler space, viz.,
\begin{equation}
V^{\,\nu}(\mathrm x)=\overline{V}^{\,\alpha}(\bar\mathrm x)\left( 
\frac{\partial\,\mathrm x^{\,\nu}}{\partial\,\bar\mathrm x^{\,\alpha}}\right) 
\label{contravariant_transformation_law} 
\end{equation}
Consider for example the complete orthogonal set of the mass-less vector
normal modes in the 2-dimensional Minkowski space $ \mathfrak{M}_{\,1+1} $
\begin{equation}
\bar \varphi_{A,\,k}^{\,\alpha}(t,x)=\left\lbrace 
\begin{array}{cc}
\varepsilon^{\,\alpha}(k)\,\bar \varphi_k(t,x) & {\rm for}\ A=L\\
\varepsilon_{\ast}^{\,\alpha}(k)\,\bar \varphi_k(t,x) & {\rm for}\ A=S
\end{array}\right. 
\end{equation}
where the dual pair of constant light-like polarization vectors is provided by
\begin{eqnarray}
\varepsilon^{\,\alpha}(k)=\left\lgroup
\begin{array}{c}
1\\{\rm sgn}(k)
\end{array}\right\rgroup\qquad\quad
\varepsilon_{\,\ast}^{\,\alpha}(k)={\textstyle\frac12}
\left\lgroup
\begin{array}{c}
1\\ -\,{\rm sgn}(k)
\end{array}\right\rgroup
\end{eqnarray}
while the scalar wave functions read
\begin{equation}
\bar \varphi_k(t,x)=(4\pi\omega)^{-\frac12}\,\mathrm e^{ikx-i\omega t}
\qquad\quad \omega=|k|\,,\,k\in\mathbb R
\end{equation}
The Minkowskian space vector normal modes do fulfill
\begin{equation}
 \bar g_{\,\alpha\beta}\left( \bar \varphi_{A,\,k}^{\,\alpha}\,,\,\bar \varphi_{B,\,p}^{\,\beta}\right)
 =\eta_{AB}\;\delta(k-p)\qquad\quad
 \eta_{AB}=\left\lgroup
\begin{array}{cc}
0 &1\\1&0
\end{array}
\right\rgroup
\end{equation}
where the invariant inner product is defined by
\begin{equation}
\bar g_{\,\alpha\beta}\left( \bar \varphi_{A,\,k}^{\,\alpha}\,,\,\bar \varphi_{B,\,p}^{\,\beta}\right)
\equiv \bar g_{\,\alpha\beta}\int_{-\infty}^{\infty}\mathrm dx\,\bar \varphi_{A,\,k}^{\,\alpha}(t,x)\,
i\!\!\buildrel\leftrightarrow\over\partial_{t}\!\bar \varphi_{B,\,p}^{\,\beta}(t,x)
 \end{equation} 
The complete and orthogonal set of the mass-less vector
normal modes in the right Rindler wedge R can be readily obtained from Eq.s
(\ref{zweibein_fields}) and (\ref{contravariant_transformation_law}).
To this concern it is expedient to introduce the accelerated polarization vectors
\begin{equation}
\left\lbrace 
\begin{array}{c}
\varepsilon^{\,\alpha}(k)\,\Lambda^{\beta}_{\phantom{\beta}\alpha}(\tau)\,
\mathrm X^{\,\nu}_{\beta}(\sigma)\equiv \varepsilon^{\,\nu}(k\,;\sigma,\tau)
\\
\varepsilon_{\,\ast}^{\,\alpha}(k)\,\Lambda^{\beta}_{\phantom{\beta}\alpha}(\tau)\,
\mathrm X^{\,\nu}_{\beta}(\sigma)\equiv \varepsilon_{\,\ast}^{\,\nu}(k\,;\sigma,\tau)
\end{array}
\right. \qquad\quad(\,\forall\,k\in\mathbb R\,)
\end{equation}
in such a manner that we eventually come to
\begin{equation}
\varphi_{A,\,k}^{\,\nu}(\varrho,\eta)=\left\lbrace 
\begin{array}{cc}
\varepsilon^{\,\nu}(k\,;\varrho,\eta)\,\varphi_k(\varrho,\eta) & {\rm for}\ A=L\\
\varepsilon_{\ast}^{\,\nu}(k\,;\varrho,\eta)\,\varphi_k(\varrho,\eta) & {\rm for}\ A=S
\end{array}\right. 
\end{equation}
where
\begin{eqnarray}
\varphi_k(\varrho,\eta)=(4\pi\omega)^{-\frac12}\,
\exp\{-i\omega\varrho[\,\sinh(\mathrm{a}\eta)-\mathrm{sgn}(k)\cosh(\mathrm{a}\eta)\,]\}
\end{eqnarray}
with $ k\in\mathbb R $ and $ \omega=|k| $.
\subsection*{Quantization of the lineal radiation field}
The manifestly covariant quantization of the free radiation field on a two dimensional 
Minkowskian space $ \mathfrak{M}_{1,1} $
in the non-homogeneous Lorenz gauge $ \partial_{\mu}A^{\mu}(\mathrm{x})=\xi B(\mathrm{x}) $
can be suitably performed 
according to the well known and long standing 
procedure and  formalism developed by Bleuler, Gupta, Lautrup and Nakanishi \cite{BGLN}
in the four dimensional case.
The Lagrangian is
\begin{equation}
{\mathcal L} = -\;{\textstyle\frac14}\,F^{\,\mu\nu}\,F_{\,\mu\nu}
+ A^{\mu}\,\partial_\mu\,\!B\,+ {\textstyle\frac12}\,\xi\,B^{2}
\label{Lagrangian} 
\end{equation}
where $B(t,x)$ is the auxiliary scalar field, while $\xi\in\mathbb R$ 
is the gauge parameter, so that the field equations read
\begin{eqnarray}
\begin{array}{c}
\left( \,\partial_{\,t}^{\,2}-\partial_{\,x}^{\,2}\,\right)A^\mu=(\,\xi-1\,)\,\partial^{\,\mu}B\\
\partial_t A^0+\partial_x A^1=\xi\,B\\
\left( \,\partial_{\,t}^{\,2}-\partial_{\,x}^{\,2}\,\right)B=0
\end{array}
\label{} 
\end{eqnarray}
It turns out that the following general canonical commutation relations hold true:
namely,
\begin{eqnarray}
[\,A^{\,\lambda}(x)\,,A^{\,\nu}(y)\,]=
i\,g^{\,\lambda\nu}\,D_0(x-y) + i\left(\,1-\xi\,\right)
\partial_{\,x}^{\,\lambda}\,\partial_{\,y}^{\,\nu}\,\mathfrak{E}(x-y)\\
\left[ F^{\,\lambda\rho}(x)\,,A^{\,\nu}(y)\right] \ =\
\left(\, g^{\;\nu\rho}\,i\partial^{\,\lambda}_{\,x}
- g^{\,\lambda\nu}\,i\partial^{\,\rho}_{\,x}\,\right)D_0(x-y)\\
\left[\,B(x)\,,A^{\,\nu}(y)\,\right]\ =\
i\partial_{\,x}^{\;\nu}\,D_0(x-y)
\end{eqnarray}
\begin{equation}
   [ \,F^{\,\rho\lambda}(x)\,,\,B(y)\,] =0\qquad\quad
\left[\,B(x)\,,\,B(y)\,\right]=0
\end{equation}
where the mass-less Pauli-Jordan real and odd distribution is
provided by 
\begin{eqnarray}
D_0(t,x)=\lim_{m\,\to\,0}\ D_{m}(t,x)={\textstyle\frac12}\,\theta(t^{\,2}-x^{\,2})\,\mathrm{sgn}(t)\\
iD_0(t,x)=
\int {{\rm d^2}k\over2\pi}\ \delta\left(k^{2}\right)\,{\rm sgn}(k_{0})\
\exp\{-\,ik_{0}t+ikx\}\\
\lim_{t\,\to\,0}\ D_0(t,x)=0\qquad\quad
\lim_{t\,\to\,0}\ \partial_{t}\,D_0(t,x)=\delta(x)\\
D_0(t,x)=D_0^{\ast}(t,x)=\ -\,D_0(-\,t,-\,x)
\end{eqnarray}
whereas $\mathfrak E(x)$ is named the { mass-less dipole-ghost invariant distribution},
which is defined by the property
\[
\left( \,\partial_{\,t}^{\,2}-\partial_{\,x}^{\,2}\,\right)\mathfrak E(t,x)=D_0(t,x)
\]
an integral-differential representation being given by
\begin{eqnarray}
\mathfrak E(t,x)&=&{\textstyle\frac12}
\left(\,t\,\partial_{\,t} - 1\,\right)\int_{-\infty}^{\infty}\mathrm{d}x'\,|x-x'|\, D_0(t,x')\nonumber\\
&\equiv& \mathfrak{D}\ast D_0(t,x)=
-\,\lim_{m\,\to\,0}\,{\partial\over\partial\,m^{2}}\,D_{\,m}(t,x)
\end{eqnarray}
An explicit expression can be obtained as follows.
Let us first consider the positive and negative Wightman functions,
i.e. the positive and negative parts
of the Pauli-Jordan distribution, in the two dimensional Minkowskian
space $ \mathfrak{M}_{1,1} $
\begin{eqnarray*}
D_{m}^{(\pm)}(t,x)&\equiv&
\frac{\pm 1}{2\pi i}\int\ 
\exp\{\pm\,ik\cdot\mathrm x\}\,\delta(k^2-m^2)\,\theta(k_0)\,\mathrm d^{2}k\\
&\equiv& \frac{\pm1}{2\pi i}
\int_0^\infty {\rm d}k\ \frac{\cos(kx)}{\sqrt{k^2+m^2}}\;
\exp\left\{\pm\,it\,\sqrt{k^2+m^2}\,\right\}
\end{eqnarray*}
whence it is clear that the positive and negative parts of
the Pauli-Jordan commutator are complex conjugate quantities
$$
[\,D_{m}^{(\pm)}(t,x)\,]^{*} = D_{m}^{(\mp)}(t,x)
$$
Consider now the integral for $ t>0 $
\begin{equation}
I(t,x)=
\int_{-\infty}^\infty\frac{\mathrm dk}{\sqrt{k^2+m^2}}\;
\exp\left\{ikx+it\,\sqrt{k^2+m^2}\,\right\}
\end{equation}
and change the variable $k=m\sinh\eta\,,$ so that
$\sqrt{k^2+m^2}=m\cosh\eta\,.$ Then we obtain
\[
I(t,x)=\int_{-\infty}^\infty\mathrm{d}\eta\ 
\exp\{im(t\cosh\eta+x\sinh\eta)\}
\]
Here $t>0$ so that two cases should be distinguished, i.e. $0<t<x$
and $t>x\,.$ After setting $\lambda\equiv t^{\,2}-x^{\,2}$ it is
convenient to carry out respectively the substitutions
\[
\left\lbrace
\begin{array}{cc}
t=\sqrt{-\lambda}\,\sinh\xi\,,\qquad 
x =\sqrt{-\lambda}\,\cosh\xi\,,& 0<t<x\\
t=\sqrt{\lambda}\,\cosh\xi\,,\qquad 
x =\sqrt{\lambda}\,\sinh\xi\,,& t>x
\end{array}
\qquad(\,t>0\,)
\right.
\]
in such a way that we can write 
\begin{eqnarray*}
I(t,x)&=& \theta(-\lambda)\int_{-\infty}^\infty \mathrm{d}\eta\
\exp\{im\sqrt{-\lambda}\,\sinh(\xi+\eta)\}\\
&+& \theta(\lambda)\int_{-\infty}^\infty \mathrm{d}\eta\
\exp\{im\sqrt{\lambda}\,\cosh(\xi+\eta)\}\\
&=& \theta(-\lambda)\int_{-\infty}^\infty \mathrm{d}\eta\
\exp\{im\sqrt{-\lambda}\,\sinh\eta\}\\
&+& \theta(\lambda)\int_{-\infty}^\infty \mathrm{d}\eta\
\exp\{im\sqrt{\lambda}\,\cosh\eta\}\qquad\qquad(\,t>0\,)
\end{eqnarray*}
Now we can use the integral representations of the Bessel
functions of real and imaginary arguments \cite{GR} 
that eventually yield for arbitrary $ t\in\mathbb R $ and $ \lambda=(t-x)(t+x) $ 
\[
I(t,\lambda)=2\theta(-\lambda)K_0(m\surd{-\lambda})
+\pi i\,\theta(\lambda)\left[{\rm sgn}(t)J_0(m\surd{\lambda})
+iN_0(m\surd{\lambda})\right]
\]
and thereby
\begin{eqnarray}
D_{m}^{(+)}(t,x)&=&
\frac{1}{4\pi}\Big\lbrace -2i\theta(-\lambda)\,K_0(m\surd{-\lambda})\nonumber\\
&+&\pi\,\theta(\lambda)\left[\,{\rm sgn}(t)\,J_0(m\surd{\lambda})
+iN_0(m\surd{\lambda})\,\right]\Big\rbrace\\
D_{m}^{(-)}(t,x)&=&
\frac{1}{4\pi}\Big\lbrace2i\theta(-\lambda)\,K_0(m\surd{-\lambda})\nonumber\\
&+&\pi\,\theta(\lambda)\left[\,{\rm sgn}(t)\,J_0(m\surd{\lambda})
-iN_0(m\surd{\lambda})\,\right]\Big\rbrace\\
D_{m}(t,x)&=&{\textstyle\frac12}\,\theta(\lambda)\,\mathrm{sgn}(t)\,J_0(m\surd{\lambda})
\end{eqnarray}
so that
\begin{equation}
D_0(t,x)={\textstyle\frac12}\,\theta(\lambda)\,\mathrm{sgn}(t)
\end{equation}
will the mass-less dipole-ghost in two space-time dimensions becomes
\begin{equation}
\mathfrak{E}(t,x)={\textstyle\frac14}\,\theta(\lambda)\,\mathrm{sgn}(t)\,\sqrt{\lambda}\lim_{m\rightarrow0}\;
\{J_1(m\surd{\lambda})/m\}={\textstyle\frac18}\,\theta(\lambda)\,\mathrm{sgn}(t)\,\lambda
\end{equation}
It is a simple and instructive exercise to verify the compatibility between the 
above general covariant canonical commutation
relations and the equations of motion, for $ \xi\not=0 $,
\begin{eqnarray*}
&&\left\{g_{\,\mu\nu}\left( \,\partial_{\,t}^{\,2}-\partial_{\,x}^{\,2}\,\right) 
-\ \left(\,1-\frac{1}{\xi}\,\right)\,\partial_{\,\mu}\,\partial_{\,\nu}\right\}\,A^{\,\nu} = 0\\
&&\left( \,\partial_{\,t}^{\,2}-\partial_{\,x}^{\,2}\,\right)B = 0\\
&&\partial \cdot A  = \xi\,B
\end{eqnarray*}
together with
\begin{equation}
\begin{array}{c}
\left( \,\partial_{\,t}^{\,2}-\partial_{\,x}^{\,2}\,\right) A^{\mu}+\partial^{\,\mu}B=0\\
\partial \cdot A=0\\
\left( \,\partial_{\,t}^{\,2}-\partial_{\,x}^{\,2}\,\right)B = 0
\end{array}
\qquad\quad(\,\xi=0\,)
\end{equation}
Moreover one can readily check that the initial conditions fulfilled by the
above general covariant canonical commutation relations are
\[
\left[\,A^{1}(t,x)\,,F(t,y)\,\right]=
-\,i\delta(x-y)=\left[\,A^{0}(t,x)\,,B(t,y)\,\right]
\]
all the remaining equal-time commutation relations being equal to zero.
The most general solution of the canonical commutation relations
in the non-homogeneous Lorenz gauge can be written in the form
\begin{eqnarray}
A^\nu(t,x)&=&\sum_{A\,=\,L,S}\int_{-\infty}^\infty\mathrm dk\;
\left[ \,f_{A,\,k}\,\bar \varphi_{A,\,k}^{\,\nu}(t,x)+
f^{\,\dagger}_{A,\,k}\,\bar \varphi_{A,\,k}^{\,\nu\,\ast}(t,x)\,\right]
\nonumber\\
&-&(\,1-\xi\,)\,\partial^{\,\nu}\,
\mathfrak D\ast B(t,x)
\label{1quantum} \\
B(t,x)&=&i\int_{-\infty}^\infty\mathrm dk\,\omega
\left[ \,f^{\,\dagger}_{S,\,k}\,\bar \varphi^{\,\ast}_{\,k}(t,x)-
f_{S,\,k}\,\bar \varphi_{\,k}(t,x)\,\right]
\label{2quantum} 
\end{eqnarray}
where $ f_{A,\,k}\ (\,A=L,S\,) $ are the destruction operators 
which satisfy the canonical commutation relations
\begin{equation}
\left[\,f_{A,\,k}\,,\,f^{\,\dagger}_{B,\,p}\,\right]
= \delta(\,p-k\,)\,\eta_{\,AB}
\label{CCRs} 
\end{equation}
all the other commutators vanishing. The canonical commutation relations
indicates that the Fock space is of indefinite metric, so that a physical Hilbert
sub-space with semi-definite metric is selected by the subsidiary condition
\begin{equation}
B^{\,(-)}(t,x)\,|\,{\rm phys}\,\rangle=0\qquad
\Longleftrightarrow\qquad
f_{S,\,k}\,\,|\,{\rm phys}\,\rangle=0
\label{subcon}
\end{equation}
where
\[iB^{\,(-)}(t,x)\equiv\int_{-\infty}^\infty\mathrm dk\;|k|\,f_{S,\,k}\,\bar \varphi_{\,k}(t,x)\]
is the positive frequency part of the auxiliary scalar field operator.
It follows therefrom that, for instance, the 1-particle states $ f^{\,\dagger}_{L,\,k}\,\,|\,0\,\rangle $
describe Lorenz lineal \textsf{nonphysical} photons, while the
1-particle states $ f^{\,\dagger}_{S,\,k}\,\,|\,0\,\rangle $
do represent the \textsf{physical} lineal scalar photons, which are all of zero norm and
satisfy (\ref{subcon}) just owing to the canonical commutation relations
(\ref{CCRs}). 

Things greatly and neatly simplify in the Feynman gauge $\xi=1$ that we shall select in what follows
without loss of generality.
Notice that if we set
\begin{eqnarray}
A^\pm=A^0\pm A^1=A_{\,\mp}
\end{eqnarray}
we can recast the covariant wave equations and the non-homogeneous Lorenz condition in the light-front form
\begin{eqnarray}
\partial_-\partial_+  A^\pm(x^-,x^+)=0\\
\partial_- A^- +\partial_+ A^+=B\\
\partial_-\partial_+ B(x^-,x^+)=0
\end{eqnarray}
It turns out that the mass-less vector wave equations and gauge condition  
in a two dimensional Minkowskian space possess the set of light-like solutions
\begin{eqnarray}
\bar \varphi_{k,L}^\nu(t,x)=\varepsilon^{\,\nu}(k)\,\bar \varphi_k(t,x)\qquad\quad
\varepsilon^{\,\nu}(k)=(1,{\rm sgn}(k))\\
\bar \varphi_{k,S}^\nu(t,x)=\varepsilon_{\ast}^{\,\nu}(k)\,\bar \varphi_k(t,x)\qquad\quad
\varepsilon_{\ast}^{\,\nu}(k)={\textstyle\frac12}\,(1,-\,{\rm sgn}(k))
\label{vectornormalmodes} 
\end{eqnarray}
where ${\rm sgn}(k)=\theta(k)-\theta(-k)$ while
\begin{equation}
\bar \varphi_k(t,x)=(4\pi\omega)^{-\frac12}\,\mathrm e^{ikx-i\omega t}
\qquad\quad \omega=|k|\,,\,k\in\mathbb R
\end{equation}
does represent the standard orthogonal set of mass-less scalar modes.
As we have
\begin{equation}
\varepsilon^{\,\mu}(k)\,\varepsilon_\mu(k)=0\qquad\quad
k_{\mu}\,\varepsilon^{\,\mu}(k)=0\qquad\quad
k^{\,\mu}=(\omega,k)
\end{equation}
it follows that the Lorenz condition is satisfied only for the longitudinal normal modes, viz.,
\begin{equation}
\partial_\nu\bar \varphi_{k,L}^{\,\nu}(t,x)=0\qquad\quad
i\partial_\nu\bar \varphi_{k,S}^{\,\nu}(t,x)=\omega\,\bar \varphi_k(t,x)
\end{equation}
Notice that the polarization vector $ \varepsilon^{\,\mu}(k) $ is indeed a
Minkowski bi-vector, because it can be written in the form
\begin{equation}
\varepsilon^{\,\mu}(k)=k^{\,\mu}\;\sqrt{\frac{2}{k\cdot k_{\,\ast}}}\quad\quad
k^{\,\mu}_{\,\ast}=(\omega,-\,k)\quad\quad
k^{\,\mu}_{\,\ast}\,k_{\mu}=k\cdot k_{\,\ast}=2|k|^{2}
\end{equation}
the other light-like  bi-vector $ k^{\,\mu}_{\,\ast} $ being the dual of the 
 bi-vector $ k^{\,\mu}=(\omega,k)$ labeling the energy-momentum carried by the normal mode
\[k^2=k_{\,\ast}^2=\varepsilon^{\,2}(k)=k\cdot\varepsilon(k)=0\qquad\quad
k_{\,\ast}\cdot\varepsilon(k)=2\omega\]
In order to set up a complete orthogonal basis of mass-less vector modes on 
the Minkowskian space $\mathfrak{M}_{1,1}\,,$ one  has to introduce the
dual light-like polarization vector $ \varepsilon_{\ast}^{\,\nu}(k)=\frac12\,(1,-\,{\rm sgn}k) $
such that
\begin{eqnarray}
\varepsilon^{\,2}_{\ast}(k)=0
\qquad\quad
\varepsilon_{\ast}(k)\cdot \varepsilon(k)=1
\qquad\quad k\cdot \varepsilon_{\ast}(k)=\omega
\end{eqnarray}
It turns out that all the normal modes (\ref{vectornormalmodes}) are of 
positive frequencies with respect to the time-like
Killing vector $ i\partial_t\,, $ for they fulfill
\begin{equation}
i\partial_{t}\bar \varphi_{k,A}^{\,\nu}=\omega\,\bar \varphi_{k,A}^{\,\nu}
\qquad\quad(\,\omega>0\,,\,A=L,S\,)
\end{equation}
It is also useful to introduce the light-front components of 
the light-like polarization vectors that read
\begin{equation}
\varepsilon^\pm(k)=\varepsilon^0\pm \varepsilon^1=2\theta(\pm k)
\qquad\quad
\varepsilon^{\,\pm}_{\,\ast}(k)=\varepsilon^{\,0}_{\,\ast}\pm\varepsilon^{\,1}_{\,\ast}
={\theta(\mp\,k)}
\end{equation}
The normal modes with $k>0$ consist of the trigonometric functions of the light-front
spatial variables $(-\,u)=x^{-}=x_{+}\,,$ see equation (\ref{uv}) 
\[(4\pi\omega)^{-\frac12}\,\varepsilon^+\,\mathrm e^{-\,i\omega x_+}
=2\theta(k)\,(4\pi k)^{-\frac12}\,\mathrm e^{-\,ik x_{+}}\equiv\varphi_{k,L}^{\,\rightarrow}(-\,u)\] 
\[(4\pi\omega)^{-\frac12}\,\varepsilon_{\ast}^-\,\mathrm e^{-\,i\omega x^-}
=\theta(k)\,(4\pi k)^{-\frac12}\,\mathrm e^{-\,ik x^-}\equiv
\varphi_{k,S}^{\,\rightarrow}(-\,u)\] 
It turns out that in the standard instant-form coordinates $ (t,x) $ those normal modes do represent
wave fronts propagating from left to right along the $ Ox- $line with unit velocity, so that
we can conclude that the trigonometric functions of the variable $ u=-\,x^{\,-}=-\,x_{+} $
do describe progressive waves.

Conversely, for $k<0$ one has to deal with the trigonometric functions of the other 
light-front temporal variables $ v=x^{\,+}=x_- $ so that
\[(4\pi\omega)^{-\frac12}\,\varepsilon^-\,\mathrm e^{\,i\omega x_-}
=2\theta(-\,k)\,(-\,4\pi k)^{-\frac12}\,\mathrm e^{\,ik x_{-}}\equiv\varphi_{k,L}^{\,\leftarrow}(v)\]
\[(4\pi\omega)^{-\frac12}\,\varepsilon_{\ast}^+\,\mathrm e^{\,i\omega x^+}
=\theta(-\,k)\,(-\,4\pi k)^{-\frac12}\,\mathrm e^{\,ik x^+}
\equiv\varphi_{k,S}^{\,\leftarrow}(v)\]
which shows that the trigonometric functions of the temporal light-front variable
$ v=x^{\,+}=x_- $ actually correspond, in instant-form
coordinates, to regressive waves moving with unit velocity from right to left
along the $ Ox- $straight line.  

It is worthwhile to remark that the mass-less normal modes $ \bar \varphi_k(t,x) $
for the auxiliary scalar field $ B(t,x) $ are normalized according to the Lorentz invariant inner product
\begin{equation}
\left( \bar \varphi_{k^{\,\prime}}\,,\,\bar \varphi_k\right) =
\int_{-\infty}^{\infty}\mathrm dx\,\bar \varphi_{k^{\,\prime}}(t,x)\,
i\!\!\buildrel\leftrightarrow\over\partial_{t}\!\bar \varphi_k(t,x)=
\delta(k-k^{\,\prime})
\end{equation}
while the polarized vector normal modes $ \bar \varphi_{k,A}^{\,\nu}(t,x) $
in the Feynman gauge are normalized
according to the Lorentz invariant inner product
\begin{equation}
\bar \varphi_{A,\,k}^{\,\nu}(t,x)=\left\lbrace 
\begin{array}{cc}
\varepsilon^{\,\nu}(k)\,\bar \varphi_k(t,x) & {\rm for}\ A=L\\
\varepsilon_{\ast}^{\,\nu}(k)\,\bar \varphi_k(t,x) & {\rm for}\ A=S
\end{array}\right. 
\end{equation}
which fulfill
\begin{equation}
 g_{\,\mu\nu}\left( \bar \varphi_{A,\,k}^{\,\mu}\,,\,\bar \varphi_{B,\,p}^{\,\nu}\right)
 =\eta_{AB}\;\delta(k-p)\qquad\quad
 \eta_{AB}=\left\lgroup
\begin{array}{cc}
0 &1\\1&0
\end{array}
\right\rgroup
\end{equation}
The vector normal modes $ \bar \varphi_{L,\,k}^{\,\nu}(t,x) $ do represent the wave functions
of the lineal photons with definite momentum and polarization satisfying the Lorenz condition
because
$ k_{\nu}\,\varepsilon^{\nu}(k)=0\,, $
while $ \bar \varphi_{S,\,k}^{\,\nu}(t,x) $ describe the scalar lineal photons because
\begin{equation}
\partial_{\nu}\,\bar \varphi_{L,\,k}^{\,\nu}(t,x)=0
\qquad\quad
i\partial_{\nu}\,\bar \varphi_{S,\,k}^{\,\nu}(t,x)=\omega\,\bar \varphi_k(t,x)
\end{equation}
Putting altogether we can write the light-front components of the Feynman gauge vector potential in the form
\begin{eqnarray}
A^+(x^+,x^-)&\equiv&A^+_{\,L}(x^-) + A^+_{\,S}(x^+)
\label{A+M}\\
A^+_{\,L}(x^-)&=&
2\int_{0}^\infty\mathrm dk\;(4\pi k)^{-\frac12}
\left[\,f_{L,\,k}\,\mathrm e^{\,-ikx^{-}} +  f^{\,\dagger}_{L,\,k}\,\mathrm e^{\,ikx^{-}}\,\right]\\
A^+_{\,S}(x^+)&=&
\int_{0}^\infty\mathrm dk\;(4\pi k)^{-\frac12}
\left[\,f_{S,-\,k}\,\mathrm e^{-\,ikx^{+}} +  f^{\,\dagger}_{S,-\,k}\,\mathrm e^{\,ikx^{+}}\,\right]
\end{eqnarray}
together with
\begin{eqnarray}
A^-(x^+,x^-)&\equiv& A^-_{\,L}(x^+) + A^-_{\,S}(x^-)
\label{A-M}\\
A_{\,L}^-(x^+)&=&2\int_0^\infty\mathrm dk\;(4\pi k)^{-\frac12}\,
\left[ \,f_{L,\,-k}\,\mathrm e^{-\,ikx^+}+f^\ast_{L,\,-k}\,\mathrm e^{\,ikx^+}\,\right]\\
A^-_{\,S}(x^-)&=&
\int_{0}^\infty\mathrm dk\;(4\pi k)^{-\frac12}
\left[\,f_{S,\,k}\,\mathrm e^{-\,ikx^{-}} +  f^{\,\dagger}_{S,\,k}\,\mathrm e^{\,ikx^{-}}\,\right]
\end{eqnarray}
Moreover, from equation (\ref{2quantum}) we get the normal mode expansion of the auxiliary scalar field
\begin{eqnarray}
B(t,x)&=&i\int_{-\infty}^\infty\mathrm dk\,\omega
\left[ \,f^{\,\dagger}_{S,\,k}\,\bar \varphi^{\,\ast}_{\,k}(t,x)-
f_{S,\,k}\,\bar \varphi_{\,k}(t,x)\,\right]\nonumber\\
&=&i\int_{0}^\infty\mathrm dk\,k\,(4\pi k)^{-\frac12}
\left[ \,f^{\,\dagger}_{S,\,k}\,\mathrm{e}^{\,ikx^-} -
f_{S,\,k}\,\mathrm{e}^{-\,ikx^-}\,\right]\nonumber\\
&-&i\int_{0}^\infty\mathrm dk\,k\,(4\pi k)^{-\frac12}
\left[ \,f_{S,-\,k}\,\mathrm{e}^{-\,ikx^+}
-f^{\,\dagger}_{S,-\,k}\,\mathrm{e}^{\,ikx^+} \,\right]\nonumber\\
&=&B(-\,u)+B(v)
\end{eqnarray}
It is worthwhile to remark that the above expressions manifestly satisfy
both the D'Alembert wave equation as well as the non-homogeneous Lorenz condition because
\begin{eqnarray}
\partial_-\,A_{\,L}^-(x^+)=\partial_+\,A_{\,L}^+(x^-)\equiv 0\\
\partial_-\,A_{\,S}^-(x^-)=B(-\,u)\qquad\quad
\partial_+\,A_{\,S}^+(x^+)=B(v)
\end{eqnarray}
while the electric field strength is provided by
\begin{eqnarray}
&&F(t,x)\equiv F^{10}(t,x)=F_{01}(t,x)=
-\partial_0\,A^1-\partial_1\,A^0\nonumber\\
&=&-\left(\partial_{+} + \partial_{-}\right){\textstyle\frac12}\left(A^+ - A^-\right)
-\left(\partial_{+} - \partial_{-}\right){\textstyle\frac12}\left(A^+ + A^-\right)\nonumber\\
&=&B(x^{\,-})-B(x^{\,+})=
\int_{0}^\infty\mathrm dk\,k\,(4\pi k)^{-\frac12}\,\mathrm{e}^{-\,\pi i/2}\nonumber\\
&\times&\left[ \,f^{\,\dagger}_{S,\,k}\,\mathrm{e}^{\,ikx^-} -
f_{S,\,k}\,\mathrm{e}^{-\,ikx^-} + f_{S,-\,k}\,\mathrm{e}^{-\,ikx^+}
-f^{\,\dagger}_{S,-\,k}\,\mathrm{e}^{\,ikx^+} \,\right]\nonumber\\
&\equiv& F(-\,u)+F(v)
\end{eqnarray}
\subsection*{The Bogoliubov coefficients}
It is very convenient to write
\begin{eqnarray}
A^+(x^{\,-})=\theta(x^{\,-})\,A^{+}(-\,u)\ +\ \theta(-\,x^{\,-})\,A^{+}(u)
\\
A^-(x^{\,+})=\theta(x^{\,+})\,A^{-}(v)\ +\ \theta(-\,x^{\,+})\,A^{-}(-\,v)
\end{eqnarray}
in such a manner that one gets a representation for the solutions of the wave equations and the Feynman
gauge subsidiary condition
in the Rindler's regions R and  L, the two regions being interchanged by parity
and time reversal symmetry. Actually, in the Rindler's regions L and R one may adopt
an alternative expansion based upon the Rindler's counterparts
$ \varphi_k $ of the massless scalar normal modes $ \bar\varphi_k $ on the Minkowskian space
$ \mathfrak{M}_{1,1}$.
To this concern, we remark that
the metric (\ref{2metrica'}) is conformal to the whole Minkowskian space, so that
under the conformal transformation 
$ g_{\mu\nu} \longmapsto \mathrm e^{-2\mathrm a\xi} g_{\mu\nu}\,,$
the line element reduces to $ \mathrm d\eta^2-\mathrm d\xi^2\,. $
Since the D'Alembert wave equation as well as the Lorenz condition are 
conformally invariant, we can recast the latter in Rindler's coordinates as
\begin{eqnarray}
\left(\,\partial_\eta^2-\partial_\xi^2\,\right)A^\pm(\eta,\xi)=0\\
\left(\,\partial_\eta\pm \partial_\xi\,\right)A^\pm(\eta,\xi)=0
\end{eqnarray}
for which there exist normal mode solutions 
\begin{equation}
(4\pi\omega)^{-\frac12}\,\exp\{ik\xi\pm i\omega\eta\}
\qquad\omega=|k|>0\,,\,k\in\mathbb R
\end{equation}
It follows that from eq.~(\ref{uv}) we eventually find for $ u,v\in\mathbb R^+\,,$ 
i.e. in the right Rindler wedge
$ \mathfrak W_R \,,$
\begin{eqnarray}
A^+(u)=\int_0^\infty\mathrm d\omega\;(4\pi \omega)^{-\frac12}\,
\left[ \,g_{L,\,\omega}\,(\mathrm a u)^{\,i\omega/\rm a} + 
g^\ast_{L,\,\omega}\,(\mathrm a u)^{-\,i\omega/\rm a}\,\right]
\label{A+R} \\
A^-(v)=\int_0^\infty\mathrm d\omega\;(4\pi\omega)^{-\frac12}\,
\left[ \,g_{L,\,-\omega}\,(\mathrm a v)^{-\,i\omega/\rm a} +
g^\ast_{L,\,-\omega}\,(\mathrm av)^{\,i\omega/\rm a}\,\right]
\label{A-R} 
\end{eqnarray}
where $ g_{L,\,\omega}\,,\,g_{L,\,-\omega} $ do represent the holomorphic amplitudes
of the Lorenz gauge potential on the the right Rindler wedge
$ \mathfrak W_R \,.$

The relation between the normal modes
expansions of the quantum fields in the Minkowski and Rindler spaces is 
well known \cite{birreldavies,unruh,fulling} and expressed by the
Bogoliubov coefficients, which satisfy a set of consistency conditions
just provided by the canonical commutation relations. It turns out that
the connection between the expansions of the Lorenz gauge potential in terms
of the Minkowski modes (\ref{A+M}-\ref{A-M}) and of the Rindler modes
(\ref{A+R}-\ref{A-R})  can be neatly obtained according to the method recently 
developed by Aref'eva and Volovich
\cite{irina}. Actually, for any real tempered distribution $ \mathrm T\in\mathsf{S}'(\mathbb R) $
the Mellin transform of its restriction $ \mathrm T_+ $ to the real positive half-line
$ v>0 $ is defined by
\begin{equation}
F_+(s)=\int_0^\infty\mathrm dv\;\mathrm T_+(v)\,v^{\,s-1}\qquad\quad\Re\mathrm e\,s>0
\end{equation}
which admits analytic continuation to a meromorphic function in the whole complex plane with 
simple poles at $ s=0,-1,-2,-3,\ldots\,. $ The inversion formula reads
\begin{eqnarray}
\mathrm T_+(v)&=&\frac{v^{-\lambda}}{2\pi}\int_{-\infty}^\infty\mathrm d\sigma\;
F_+(\lambda+i\sigma)\,v^{-i\sigma}\nonumber\\
&=&\frac{v^{-\lambda}}{2\pi}\int_0^\infty\mathrm d\sigma\;
F_+(\lambda+i\sigma)\,v^{-i\sigma}\ +\ \rm c.c.
\end{eqnarray}
where $v$ and $\lambda$ are real positive numbers. A comparison with eq. (\ref{A-R})
yields $ \sigma=\omega/\rm a $ and $ \lambda\rightarrow0\,, $
in such a manner that for $ v>0\,,\,\sigma>0 $
\begin{equation}
\mathrm T_+(v)\equiv A_+(v)\qquad\quad
F_+(i\sigma)=\mathrm a^{\frac12-i\sigma}\;\sqrt{\frac{\pi}{\sigma}}\;g_{L,\,-\sigma\rm a}
\end{equation}
Moreover we get
\begin{eqnarray}
&&\int_0^\infty{\mathrm dv}\;A^-(v)\,v^{\,s-1}\nonumber\\
&=&\int_0^\infty\mathrm dv\;v^{\,s-1}
\int_0^\infty\mathrm dk\;(4\pi k)^{-\frac12}\,
\left[ \,f_{L,\,-k}\,\mathrm e^{-\,ikv}+f^\ast_{L,\,-k}\,\mathrm e^{\,ikv}\,\right]
\end{eqnarray}
where use has been made of the Minkowski set (\ref{A-M}). From the relation
\[\lim_{\epsilon\,\rightarrow\,0^+}\int_0^\infty{\mathrm dv}\;\exp\{\,\pm\,i(k\pm i\epsilon)\}\,v^{\,s-1}
=\Gamma(s)\,k^{-s}\,\mathrm e^{\,\pm\pi is/2}\]
we definitely obtain
\begin{eqnarray}
F_+(s)&=&
{\Gamma(s)\over2\surd\pi}\int_0^\infty\mathrm dk\;k^{-s-\frac12}\,
\left[ \,f_{L,\,-k}\,\mathrm e^{\,-\pi is/2}+f^\ast_{L,\,-k}\,\mathrm e^{\,\pi is/2}\,\right]
\nonumber\\
F_+(i\sigma)&=&{\Gamma(i\sigma)\over2\surd\pi}\int_0^\infty\mathrm dk\;k^{-i\sigma-\frac12}\,
\left[ \,f_{L,\,-k}\,\mathrm e^{\,\pi\sigma/2}+f^\ast_{L,\,-k}\,\mathrm e^{\,-\pi\sigma/2}\,\right]\nonumber\\
&=&\mathrm a^{\frac12-i\sigma}\;\sqrt{\frac{\pi}{\sigma}}\;g_{L,\,-\sigma\rm a}
\end{eqnarray}
that eventually yields the relationships between the holomorphic amplitudes of the
Lorenz gauge potential on the Minkowskian space $ \mathfrak{M}_{1,1} $ and 
the right Rindler wedge $ \mathfrak{W}_{R} $ respectively: namely,
\begin{eqnarray}
g_{L,\,-\omega}&=&{\sqrt{\omega}\over2\pi\mathrm a}\;
\Gamma\left({i\omega\over\rm a}\right)
\nonumber\\
&\times&\int_0^\infty{\mathrm dk\over\surd k}\;\left({k\over\rm a}\right)^{-i\omega/\rm a}\,
\left[ \,f_{L,\,-k}\,\mathrm e^{\,\pi\omega/2\rm a} +
f^\ast_{L,\,-k}\,\mathrm e^{\,-\pi\omega/2\rm a}\,\right]
\label{bogo-} 
\end{eqnarray}
in full accordance with \cite{irina}. The above operator transformation (\ref{bogo-})
leads to the coefficients
\begin{eqnarray}
\alpha_{-k,-\omega}=
\sqrt{{\omega\over2\pi\mathrm a}}\;
\Gamma\left(-\,{i\omega\over\rm a}\right)
\left({k\over\rm a}\right)^{\,i\omega/\rm a}\,
\mathrm e^{\,\pi\omega/2\rm a}\\
\beta_{-k,-\omega}=
\sqrt{{\omega\over2\pi\mathrm a}}\;
\Gamma\left({i\omega\over\rm a}\right)
\left({k\over\rm a}\right)^{-i\omega/\rm a}\,
\mathrm e^{-\pi\omega/2\rm a}
\end{eqnarray}
that yields
\begin{eqnarray}
|\,\alpha_{-k,-\omega}\,|^2\,=\,
\frac{\mathrm e^{\,2\pi\omega/\rm a}}{\mathrm e^{\,2\pi\omega/\rm a}-1}
=1-N_{\,\omega,T}\\
|\,\beta_{-k,-\omega}\,|^2\,=\,
\frac{1}{\mathrm e^{\,2\pi\omega/\rm a}-1}
\equiv{N}_{\,\omega,T}
\end{eqnarray}
which actually corresponds to  Bose-Einstein thermal distributions  at the equilibrium temperature
$T=\hbar\mathrm a/2\pi ck_{\,\rm B}\,,$ i.e. the Unruh temperature \cite{unruh}, $ k_{\,\rm B} $
being the Boltzmann constant.
A quite similar calculation drives to
\begin{eqnarray}
&&\int_0^\infty{\mathrm du}\;A^+(u)\,u^{\,s-1}\nonumber\\
&=&\int_0^\infty\mathrm du\;u^{\,s-1}
\int_0^\infty\mathrm dk\;(4\pi k)^{-\frac12}\,
\left[ \,f_{L,\,k}\,\mathrm e^{-\,iku}+f^\ast_{L,\,k}\,\mathrm e^{\,iku}\,\right]
\end{eqnarray}
where use has been made of the Minkowski set (\ref{A+M}). Hence we definitely obtain
\begin{eqnarray}
F^+(s)&=&
{\Gamma(s)\over2\surd\pi}\int_0^\infty\mathrm dk\;k^{-s-\frac12}\,
\left[ \,f_{L,\,k}\,\mathrm e^{\,-\pi is/2}+f^\ast_{L,\,k}\,\mathrm e^{\,\pi is/2}\,\right]\\
F^+(i\sigma)&=&\mathrm a^{\frac12-i\sigma}\;
\sqrt{\frac{\pi}{\sigma}}\;g^{\ast}_{L,\,\sigma\rm a}\nonumber\\
&=&{\Gamma(i\sigma)\over2\surd\pi}\int_0^\infty\mathrm dk\;k^{-i\sigma-\frac12}\,
\left[ \,f_{L,\,k}\,\mathrm e^{\,\pi\sigma/2}+f^\ast_{L,\,k}\,\mathrm e^{\,-\pi\sigma/2}\,\right]
\end{eqnarray}
and thereby
\begin{eqnarray}
g_{L,\,\omega}^{\ast}&=&{\sqrt{\omega}\over2\pi\mathrm a}\;
\Gamma\left({i\omega\over\rm a}\right)
\nonumber\\
&\times&\int_0^\infty{\mathrm dk\over\surd k}\;\left({k\over\rm a}\right)^{-i\omega/\rm a}\,
\left[ \,f_{L,\,k}\,\mathrm e^{\,\pi\omega/2\rm a} +
f^\ast_{L,\,k}\,\mathrm e^{\,-\pi\omega/2\rm a}\,\right]\\
\alpha_{k,\,\omega}&=&
\sqrt{{\omega\over2\pi\mathrm a}}\;
\Gamma\left({i\omega\over\rm a}\right)
\left({k\over\rm a}\right)^{-i\omega/\rm a}\,
\mathrm e^{-\pi\omega/2\rm a}=\beta_{-k,-\omega}
\end{eqnarray}
\subsection*{Conclusion}
In this short note we have presented the quantum theory of the lineal, i.e. one dimensional in space,
radiation field in a uniformly accelerated reference frame referred to Rindler's curved coordinates.
The pair of physical and nonphysical radiation fields, which must be
introduced in a diffeomorphism and gauge invariant quantum theory on a flat space-time, 
appear to be described by null norm quantum fields, in such a manner that a subsidiary condition must
be introduced to select the physical Hilbert sub-space, the quantum states of which are of positive
semi-definite norm according to \cite{BGLN}. The Bogoliubov coefficients connecting the inertial and
non-inertial Observers have been calculated by means of a new technique due to Aref'eva and Volovich
\cite{irina}. The result of our calculations turns out to be 
singularity free and in full agreement with the long standing known 
expression \cite{unruh}, which is valid for an ordinary mass-less scalar field. This conclusion allows 
to set upon a solid and reliable framework the operational analysis of \cite{hawton} concerning the
emission, propagation and detection of the electromagnetic radiation in a non-inertial reference frame.

\begin{thebibliography}{2015}
%
\bibitem{birreldavies}
N.D. Birrel and P.C.W. Davies (1982)
\newblock {\em Quantum Fields in Curved Space},
\newblock Cambridge University Press, Cambridge (UK).

\bibitem{hawton}
Margaret Hawton (2013) \textit{Photon counting by inertial and accelerated detectors},
Phys. Rev. A \textbf{87}, 042116, 
arXiv:1304.5138v1 [quant-ph].

\bibitem{CSthesis}
Caterina Specchia (2013)\emph{Vector Fields in a Rindler Space}, AMS Tesi di Laurea,
Bologna University.

\bibitem{unruh}
W.G. Unruh (1976) \emph{Notes on black-hole evaporation}, Phys. Rev. D \textbf{14}, 870-892.

\bibitem{fulling}
S.A. Fulling (1973)
\emph{Nonuniqueness of Canonical Field Quantization in Riemannian Space-Time},
Phys. Rev. D {\bf 7}, 2850.

\bibitem{fedotov}
A.M. Fedotov, V.D. Mur, N.B. Narozhny, V.A. Belinski\u\i\
and B.M. Karnakov (1999) Phys. Lett. A {\bf 254}, 126-132.

\bibitem{lonsol11}
P. Longhi and R. Soldati (2011) \emph{The Unruh Effect Revisited},
Phys. Rev. D{\bf 83}, 107701;
arXiv:1101.5976 [hep-th].

\bibitem{castorina}
P. Castorina and M. Finocchiaro (2012) \emph{Symmetry Restoration by Acceleration},
Jour. Mod. Phys. {\bf 3}, 1703; arXiv:1207.3677 [hep-th].

\bibitem{cotaescu}
Ion I. Cotaescu (2015) \emph{Acceleration in de Sitter spacetimes},
Europhys. Lett. \textbf{109}, 4, 40002; arXiv:1407.2502 [gr-qc];
(2013) \emph{How to kill the Unruh effect},  e-Print arXiv:1301.6650v4 [gr-qc].

\bibitem{linet}
B. Linet,
\emph{Static, massive fields and vacuum polarization potential in Rindler space},
Int. J. Mod. Phys. D {\bf 7} (1998) 61-71.

\bibitem{lenz}
F. Lenz, K. Ohta and K. Yazaki (2008)
\emph{Canonical quantization of gauge fields in static space-times with application to Rindler spaces},
Phys. Rev. D {\bf 78}, 065026; arXiv: 0803.2001v3 [hep-th].

\bibitem{CHM08}
Lu\'{\i}s C.B. Crispino, Atsushi Higuchi and George E.A. Matsas (2008)
\emph{The Unruh Effect and its Applications}, Rev.  Mod.  Phys. \textbf{80}, 787-838,
arXiv: 0710.5373 [gr-qc].

\bibitem{oriti}
Daniele Oriti (2000) \emph{The Spinor Field in Rindler Space-Time: An Analysis of the Unruh Effect},
Il Nuovo Cimento \textbf{B115}, 1005-1024.

\bibitem{lonsol13}
P. Longhi and R. Soldati (2013) \emph{Neutral Massive Spin $ \frac12 $ Particle Emission in a Rindler Space},
Int. Jour. Mod. Phys. A {\bf 28}, 1350109;
arXiv:1210.7378v3 [hep-th].

\bibitem{BGLN}
K.~Bleuler, Helv. Phys. Acta {\bf 23} (1950) 567;
Sen N.~Gupta, Proc. Phys. Soc. A{\bf 63} (1950) 681;
B.~Lautrup, Kgl. Danske Videnskab. Selskab. Mat.-fys. Medd. 
{\bf 35} (1967) No. 11, 1;
Noboru Nakanishi, Prog. Theor. Phys. {\bf 35} (1966) 1111;
{\em ibid}. {\bf 49} (1973) 640; {\em ibid}. {\bf 52} (1974) 1929;
Prog. Theor. Phys. Suppl. No. {\bf 51} (1972) 1.

\bibitem{irina}
I. Ya. Aref'eva and I.V. Volovich (2013)
\emph{Note on the Unruh Effect},
e-Print arXiv:1302.6699 [hep-th].

\bibitem{rindler}
Wolfgang Rindler, \emph{Kruskal Space and the Uniformly Accelerated Frame},
 Am. J. Phys. \textbf{34} (1964) 1174;
\textit{Essential Relativity}, New York: Van Nostrand, 1969.


\bibitem{sokolovsky}
M. Sokolovsky (2013)
\emph{Rindler Space and Unruh Effect},
e-Print arXiv:1304.2833v2 [gr-qc].

\bibitem{GR}
\newblock I.S.~ Gradshteyn and I.M.~ Ryzhik, 
\newblock {\em Table of Integrals, Series, and Products}, 
\newblock Fifth Edition, Alan Jeffrey Editor, Academic Press, San Diego, 1996.

%
\end{thebibliography}
\end{document}